\documentclass[aps,showpacs,pre,twocolumn]{revtex4}

\usepackage{psfig}

\begin{document}
\title{Vibration-induced granular segregation: a phenomenon driven by 
three mechanisms}

\author{D. A. Huerta and J. C. Ruiz-Su\'arez \cite{JCRS}} 

\affiliation{CINVESTAV del IPN, Depto.\ de F\'\i sica Aplicada, \\
  97310 M\'erida, Yucat\'an, M\'exico }

\date{\today}

\begin{abstract}
The segregation of large spheres in a granular bed under vertical
vibrations is studied. In our experiments we
systematically measure rise times as a function of density, diameter and 
depth; for two different
sinusoidal excitations. The measurements reveal that: at low frequencies, 
inertia and  convection are the only mechanisms behind segregation. 
Inertia (convection) dominates
when the relative density is greater (less) than one. At high frequencies, 
where convection is suppressed, 
fluidization of the granular bed causes either buoyancy or sinkage and 
segregation occurs.
\end{abstract}


\maketitle


Many theoretical and experimental studies have been carried out in the last five decades aimed
to reveal the physics of one of the most intriguing phenomena in granular matter: vibration-induced 
segregation 
\cite{Williams63,Rippie67,Ahmad73,Cooke76,Williams76,Rosato87,Duran93,Nagel93,Muzzio,Brone,Cooke96,Vanel,Jullien,Rosato2002}. 
However, when this phenomenon intensively investigated appeared to be 
well understood, new scientific puzzles came 
into scene \cite{Nagel2001,Liffman,Hong2001,Burtally,Hong2002,Tixco,chinos,Nahmad2003}. Air-driven
segregation, inertia, condensation, are 
now new words added to the already vast list of concepts important in the subject. 
Thus, since this problem is an important concern to industries dealing with 
granulates, these recently 
disclosed effects should be further investigated. 
Granular segregation was first reported in 1939 by Brown \cite{Brown} and studied ever since by the 
engineering community \cite{Williams63,Rippie67,Ahmad73,Cooke76,Williams76}, until 
it was brought in 1987 to the physics realm  with the suggestive  
name of ``Brazil Nut Problem'' (BNP) \cite{Rosato87}.  Results related to this problem established 
themselves as benchmarks of granular segregation. But the question: why the Brazil nuts are on top, 
seems to be yet an open matter of discussion.  Both theoretical and experimental studies have focused on 
the influence of size, friction, density and excitation parameters 
\cite{Duran93,Nagel93,Muzzio,Brone,Vanel,Jullien,Liffman,Nahmad2003,libro1,libro2} 
and the results explain, or obscure,  bit by bit the underlying mechanism  behind the BNP. 
Some of these results support the idea that it is ``void-filling'' beneath large ascending particles, the mechanism 
promoting the upward movement of an intruder in a shaken granular bed \cite{Rosato87,Rosato2002}. 
Other researches claim that global convection is the driven force behind 
the BNP \cite{Nagel93}, 
and others, that arching \cite{Duran93} or inertia \cite{Liffman,Nahmad2003} are crucial elements to explain it . The 
dilemma is not 
yet settled on the verge of even more recent findings \cite{Nagel2001}; being the most relevant the 
surprising result that 
decreasing the density of the intruder does not necessarily mean a monotonic increasing of
the rise time, as might be previously suggested by studies in 3D \cite{Muzzio} and 2D 
\cite{Liffman}. Furthermore, based on computer simulations Hong et al even dared to predict
the reverse segregation effect in the BNP \cite{Hong2001,Hong2002} (known now in the literature as the 
RBNP), which was immediately confronted by two groups \cite{Tixco,Walliser}, but nevertheless observed in the 
laboratory by Breu et al \cite{germans}. Finally, Yan et al \cite{chinos} recently failed to confirm the experimental 
findings of M$\ddot{o}$bius et al \cite{Nagel2001}.  

Based on this debate, a simple, yet overwhelming conclusion 
arises: more research is needed if we want to uncover the physics of this elusive granular matter problem. 
This Letter aims to contribute to its final understanding. 

Our experimental set up consists of a Plexiglas cylinder (closed at one end) of 10 cm inner 
diameter and  26 cm of length. The Plexiglas cylinder is fixed to a vibrating table fed with an amplified
periodic voltage coming from a function generator (HP-33120A). In a typical experiment, the column is filled
with small seeds or glass beads. Rise times are measured by a stopwatch. Excellent reproducibility of the
data is obtained if the temperature and humidity do not change during the experiments.

\begin{figure}[h]
\centerline{\psfig{figure=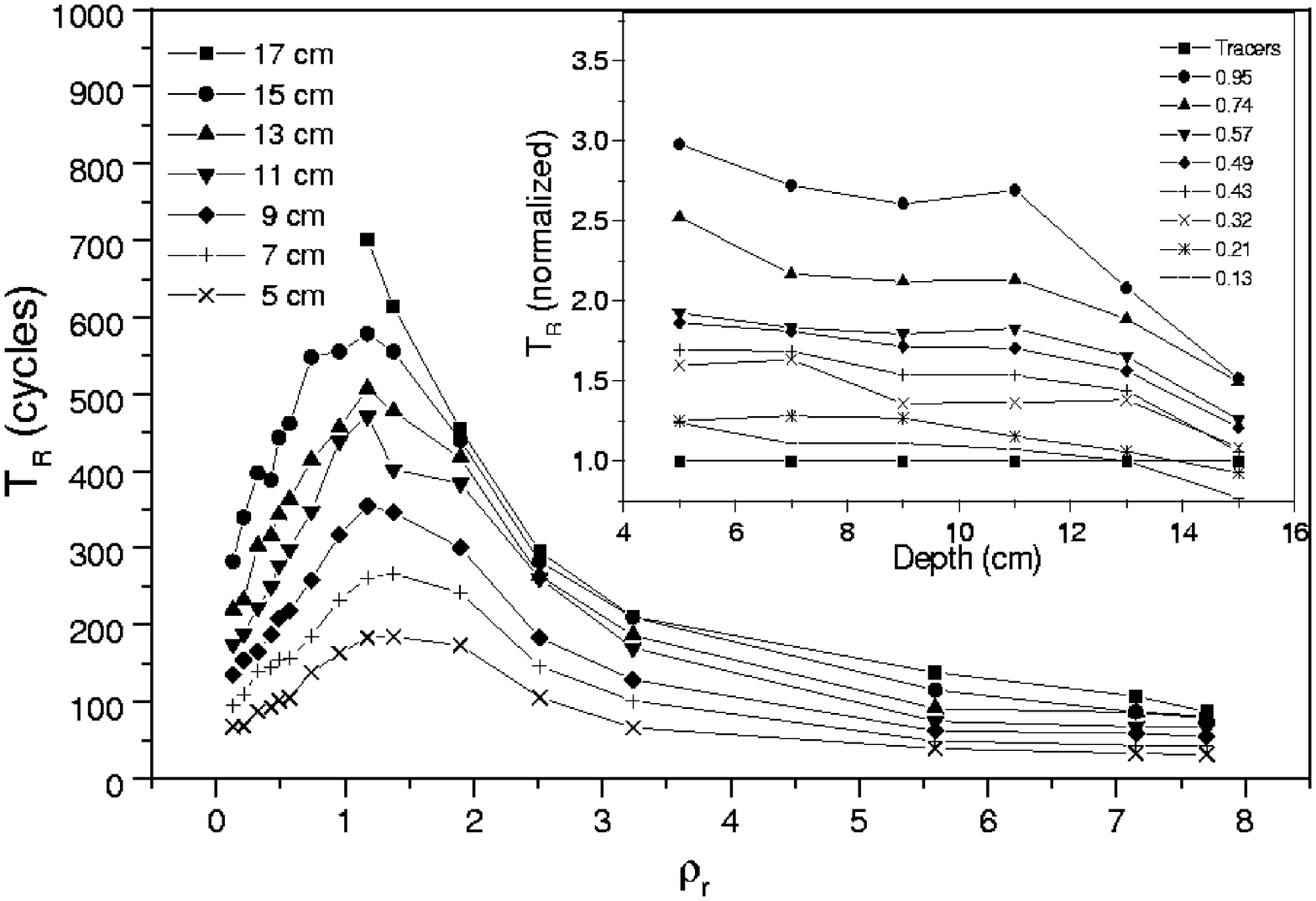,width=8.3cm,angle=270}}
\caption{{} Dimensionless rise times for intruders with $\rho_r > 1$ for all depths, normalized with 
the rise times of the fastest (densest) sphere, at each depth. The line is the best fit obtained 
with $a \rho_r ^b$, giving $a = 9.84$ and  $b = - 1.09$. Inset: dimensionless rise times for intruders 
with $\rho_r  < 1$ for 
different depths, normalized with the rise time of the fastest (lightest) 
sphere, at each depth.}
\label{densities}
\end{figure}

In Fig.\ref{densities} we show the rise 
time as a function of the intruder relative density ($\rho_r = \rho / \rho_b$,  where $\rho_b$ is the 
density of the bed particles) for 7 different starting depths: 5,7,9,11,13,15 and 17 cm below 
the surface, at 5 Hz and $\Gamma = 3$. The bed column had 21 cm of height (hence, 
at 17 cm of depth our 4 cm diameter intruders touch the bottom of the  container). The size and 
density of the bed particles we used (tapioca monodisperse spheres) are respectively 3.1 mm and 0.57 
g/cc. The intruders are plastic spheres filled with different materials to change their densities.

\begin{figure}[h]
\centerline{\psfig{figure=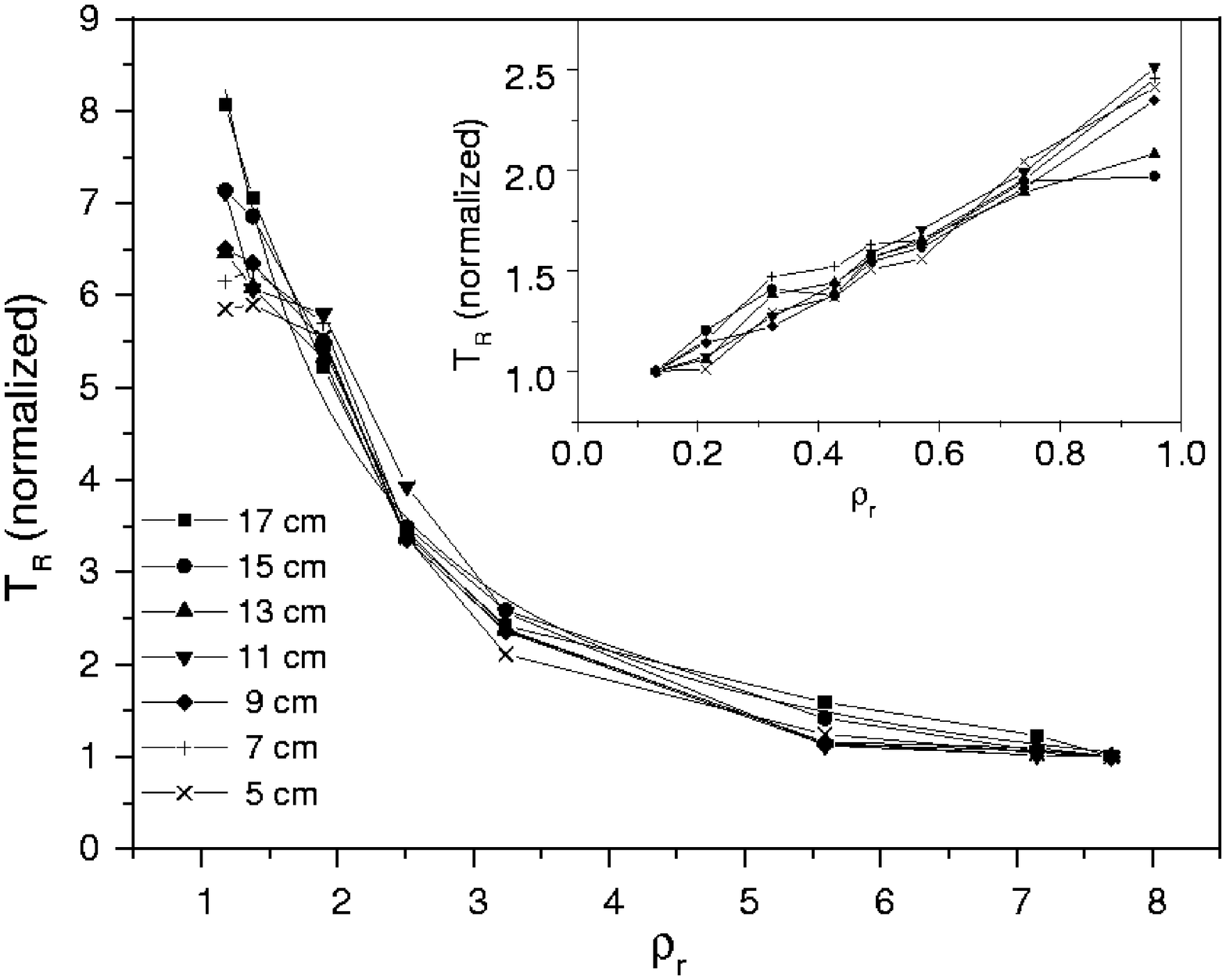,width=8cm,angle=270}}
\caption{{} Dimensionless rise times for intruders with $\rho_r > 1$ for all depths, normalized with 
the rise times of the fastest (densest) sphere, at each depth. The line is the best fit obtained 
with $a \rho_r ^b$, giving $a = 9.84$ and  $b = - 1.09$. Inset: dimensionless rise times for intruders 
with $\rho_r  < 1$ for 
different depths, normalized with the rise time of the fastest (lightest) 
sphere, at each depth. }
\label{normalized-densities1}
\end{figure}

Our first clear observation is that the ascension dynamics of the spheres whose starting positions were at the
bottom of the column, has a  monotonic dependence on
$\rho_r$; the curve diverging at $\rho_r \approx  1$.
At any other depth, the spheres regardless their density segregate to the surface following a non-monotonic
dynamics. This non-monotonic ascension dynamics was previously observed by the group of 
Chicago \cite{Nagel2001}, although the peak they observed was positioned at a relative 
density less than one and their measurements correspond to only one depth (around 5 cm).
We can normalize the data with $\rho_r > 1$ in Fig.\ref{densities} 
by  making  the rise time of 
the heaviest sphere equal to one (for each depth). In this way, the rise time of every sphere will be 
measured with the time scale of the fastest (the densest) one. These 
normalized results, 
for the spheres denser than the granulate, collapse into the
same curve, see Fig.\ref{normalized-densities1}. Based on this result we believe that these
spheres segregate mostly by inertia. 
This concept has been evoked in the literature for some time 
already
\cite{Muzzio,Liffman,Nagel2001,Burtally,Nahmad2003}, but never used to quantify granular segregation in 3D.
In a recent paper \cite{Nahmad2003}, we proposed a simple theoretical model to explain the rise 
dynamics of heavy spheres in a vibrated granulate. The model is based on energy considerations and 
states that, on each cycle,  the kinetic energy of the intruder is lost by friction during its penetration
into the granular bed; $1/2 m v_{to}^2 = \beta(h) P_l $ (where $v_{to}$  is 
the "take-off" velocity the bead has when the granulate reaches a negative  
acceleration  $a = -g$,  $m$ the mass of the bead, and $\beta(h)$ the friction force exerted upon 
it by the granulate). Therefore, since in our experiments the volume of the intruders and the vibration conditions
are constants, the penetration length per 
vibration cycle should be directly proportional to the 
density. Thus, the number of cycles for a sphere to segregate to the surface is inversely proportional to it. 
The smooth  line in Fig.\ref{normalized-densities1} was  obtained by using  $a \rho_r ^ b$, where the best fit gives 
$b = - 1.09$. What is very interesting is that all peaks in Fig.\ref{densities} are at  $\rho_r \approx  1$
and afterwards, lighter intruders start to ascend faster.  
In the inset of Fig.\ref{normalized-densities1} we show normalized rise times for $\rho_r < 1$. For this 
case, normalization was done using also the fastest sphere (the lightest). 
We will come back to this plot later.

\begin{figure}[h]
  \centerline{{\bf a)} \psfig{figure=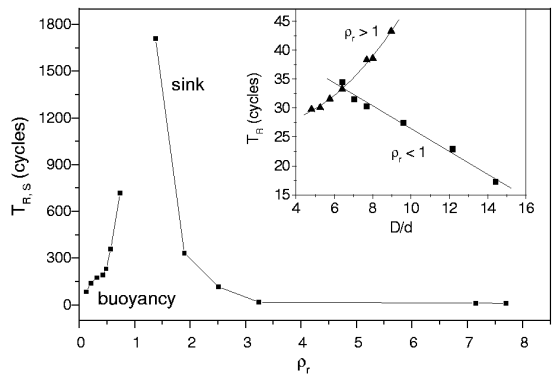,width=8.4cm,angle=270}}
\caption{{} Dimensionless rise (sink) times as a function of relative density 
for only one depth (5 cm), at 
50 Hz. In the left part of the figure ($\rho_r  < 1$) we note that intruders 
ascend faster the lighter they are, 
while in the right part ($\rho_r  > 1$) we see that the denser sink 
faster. Inset: rise times as a function
of relative diameters for $\rho_r  > 1$ (where the solid line represents 
a parabolic fit) and $\rho_r  < 1$
(where the solid line is only to help the eye). Both cases are at 
constant mass (7.1 g and 2.2 g, respectively), 5 Hz, $\Gamma = 3$, 
and 5 cm depth.}
\label{highfreq}
\end{figure}

We carried out a second experiment. The rise time of light spheres 
was measured again as we change $\rho_r$, but now the spheres are always positioned at the   
bottom of a container filled by a granulate of 6 cm of height. The frequency we use is  50 Hz
(with $\Gamma = 3$). 
We can see  that the lighter the spheres the faster they ascend, see Fig.\ref{highfreq}. 
This is contrary to what happens at low frequencies, where regardless of 
how shallow is the granular bed 
on top of the spheres, they can not segregate if they are at the bottom. 
Indeed, on one hand, at low frequencies spheres 
with $\rho_r < 1$  can not 
ascend from the bottom as seen in Fig.\ref{densities} (but spheres 
with $\rho_r > 1$ can). On the
other hand, at high frequencies spheres with $\rho_r < 1$ can ascend from 
the bottom (but spheres with $\rho_r > 1$ can not). 
During this high-frequency experiment tracer particles were put at the 
base of the container, 
alone or together with the 
light spheres, and the latter emerge but the tracer do not 
(i.e. there is no convection). 
To explain this effect, instead of "void-filling", we prefer to use a term 
used in fluids and mentioned already by some authors in the granular 
field \cite{Nagendra}: buoyancy. Is 
this granular effect caused by fluidization the missing term for 
explaining the results 
reported by the group of Chicago \cite{Nagel2001} and ours in Fig.\ref{densities}? 
The answer is no. 

The condition for buoyancy to happen, when neither convection nor inertia 
are present, is pure fluidization. 
Here, we would like to point out that fluidization with no convection is 
achieved if along the granular column there isn't a "temperature" gradient; 
were the particles of the granulate
move only around their equilibrium positions, whether or not the bed 
crystallizes \cite{Vanel}.
 
Hence, at high  frequencies the 
granulate fluidizes and buoyancy takes place, 
segregating to the top any light intruder buried inside it. 
Furthermore, at these fluidized  conditions, where light particles are 
buoyant, heavy intruders, as in common fluids, must sink. In 
fig.\ref{highfreq} we also depict sink times 
as a function of $\rho_r$ for $\rho_r > 1$. The sink curve shows that the 
heavier is the intruder, the 
faster it sinks  (we plot the time it takes for each sphere to sink its 
own diameter). 
The above concepts explains the RBNP predicted  by 
Hong et al \cite{Hong2001,Hong2002} and 
later observed by Breu  et al\cite{germans}. It also explains why 
Canul-Chay  et al \cite{Tixco} 
could not observe it in their own experiments \cite{note1}. 

To understand  the still unexplained left part of 
the M$\ddot{o}$bius et al curve and ours  in 
Fig.\ref{densities}, let us explore carefully 
the "convection connection", postulated by the group 
of Chicago \cite{Nagel93}. We can plot the same data already plotted 
in Fig.\ref{densities} in a 
different fashion, see the inset of the figure. Normalized rise 
times are plotted as 
function of depth for all relative densities less than one. These data were 
normalized using  rise 
times of tracer particles, measured at the same excitation conditions 
for each depth, while no intruders were in the bed. The 
horizontal line corresponds precisely to the tracer particles rise 
times normalized to themselves. We note 
that most of the curves are above the tracer line, indicating that 
convection is faster than the ascension of these spheres. We suggest 
then that it is pure 
convection (neither inertia because particles are light nor buoyancy 
because the bed is not 
fluidized at these low frequencies) the mechanism needed to explain the 
drop of rise time 
curves at $\rho_r \approx 1$. Looking again to the inset 
of Fig.\ref{normalized-densities1}, we
see that the normalized rise times reasonably collapse into the same curve. 
The fact that they collapse does not mean that the spheres rise at the 
tracer times, as one can clearly see
in the inset of Fig.\ref{densities}. Obviously, although convection is 
behind the ascension, their inertia is still non-negligible. Therefore, the 
heavier they are, the harder to be dragged to the top by the streaming 
flux or convective cell.
This can be understood if we look at the transfer of momentum; where both, 
the "streaming" and "collisional" modes participate in the transport of 
intruders having not-negligible masses \cite{Nagendra}.

Our results plotted in Fig.\ref{densities} 
seem to be in 
contradiction with prior experimental studies conducted by 
Shinbrot et al \cite{Muzzio} and Yan et al \cite{chinos}. 
On one hand, Shinbrot et al measure convection periods where divergence 
is seen for light intruders. On the other, 
Yan et al observe that the lighter the intruders the faster they sink. 
In both cases, the differences with our
experiments lie on the following two issues: a) in 
our experiments we always positioned the intruders at specific depths 
and measure rising times, instead of convection periods \cite{Muzzio}
or sinking times \cite{chinos}, b) the size of the bed particles.
We are convinced that if Shinbrot et al had positioned their intruders at 
specific depths in a granular bed with larger particles, they would 
have observed the non-monotonic rise dynamics we observe in 
Fig.\ref{densities}. Furthermore, Yan et al suggested that
in order to understand the non-monotonic segregation dynamics 
shown in Fig.\ref{densities}, we need to understand first 
the role of the air pressure gradient acting on the granular 
bed \cite{chinos}. According to our results, we conclude that
this suggestion is not correct. Air is indeed an important ingredient 
only in granulates with very small particles where 
reverse \cite{Muzzio} or negative \cite{chinos} buoyancy is observed.

Finally, let us remind the reader that all theoretical and experimental 
studies 
published so far on granular segregation agree on the following: 
the larger the intruder the 
faster it rises through a vibrating bed. Moreover, some geometrical 
approaches conclude that there is a critical diameter below which 
intruders do not ascend \cite{Duran93,Jullien}. The above 
conclusions are only true when the density remains constant as one 
changes the size of the intruders (see for instance, the recent 
work reported by the group of Chicago \cite{Nagel2001}). Under these 
circumstances, the mass varies with the cube of the diameter and 
therefore, due to their greater value, larger intruders ascend faster.
However, the opposite behavior is observed when the mass (not the 
density) remains constant as one changes the size of the 
intruders, see the inset of Fig.\ref{highfreq}). In this case, the faster 
intruders are the smaller (in contradiction with the mentioned 
size paradigm). Our above model, that predicts 
the hyperbolic dynamics for intruders denser than the 
granulate (at constant intruder volume, see Fig.\ref{normalized-densities1}), 
predicts also the parabolic behavior found for size segregation (see 
the inset of Fig.\ref{highfreq}) at constant 
mass with $\rho_r > 1$ \cite{note2}. In the inset of Fig.\ref{highfreq}, 
we show  rising times versus diameter for $\rho_r < 1$
(constant mass), verifying in this case what granular scientists have 
concluded in the past: larger intruders rise faster.

We report experimental results that shed definitive light to explain the 
fundamental aspects of the 
fascinating BNP. We conclude that there are only three physical mechanisms 
behind the segregation  of a large or small, heavy or light, particle in 
a vibrated granulate:
inertia, convection and buoyancy (sinkage). The first two are always 
present at $\Gamma > 1$
and low frequencies, where a non-monotonic ascension dynamics is observed 
as one changes the relative density of the intruders. Inertia dominates when 
$\rho_r > 1$, and convection does it when  $\rho_r <  1$. Segregation, 
by buoyancy or sinkage, is present at excitation conditions where the 
granulate is fluidized with no convection ($\Gamma > 1$, small amplitudes 
and high frequencies). Finally, when intruders have the same mass but 
diameters are varied keeping their relative densities greater (less) than 
one, the smaller (larger) they are, the faster they segregate to the 
free surface of the bed.




Useful discussions with C. Vargas, V. Sosa and Y. Nahamd 
are acknowledged. Special thanks are given to C. Moukarzel who suggested
the size segregation experiment keeping the mass of the intruders 
constant.
This work has been partially supported by Conacyt, M\'exico, under 
Grant No. 36256E.


\begin{thebibliography}{*}
\bibitem{note2} In the expression: $1/2 m v_{to}^2 = \beta(h) P_l $, the 
left part is constant.
Since $\beta(h)$ is proportional to the intruder cross 
section ($D^2$), $T_R$ is also proportional to it.
\end{thebibliography}
\end{document}